\documentclass[12pt,preprint]{aastex}
\usepackage{xcolor}

\def\beq{\begin{equation}}
\def\eeq{\end{equation}}
\def\ben{\begin{eqnarray}}
\def\een{\end{eqnarray}}
\def\lcdm{\Lambda{\rm CDM}}
\def\js{\hat{\bf j}_{\star}}
\def\jd{\hat{\bf j}_{d}}
\def\jz{\hat{\bf j}_{\rm ini}}

\def\rv{r_{\rm v}}

\def\rh{R_{1/2\star}}
\def\mg{\tilde{M}_{g}}
\def\munit{\,h^{-1}\,M_{\odot}}
\def\dunit{\,h^{-1}{\rm Mpc}}
\def\tm{\hat{\bf t}_{1}}
\def\ti{\hat{\bf t}_{2}}
\def\tn{\hat{\bf t}_{3}}
\def\rf{R_{f}}

\begin{document}
\title{Reoriented Memory of Galaxy Spins for the Early Universe}
\author{Jun-Sung Moon$^{1,2}$ and Jounghun Lee$^{1}$}
\affil{$^1$Astronomy Program, Department of Physics and Astronomy,
Seoul National University, Seoul 08826, Republic of Korea
\email{jsmoon.astro@gmail.com, cosmos.hun@gmail.com}}
\affil{$^2$Research Institute of Basic Sciences, Seoul National University, Seoul 08826, Republic of Korea}
\begin{abstract}
Galaxy spins are believed to retain the initially acquired tendency of being aligned with the intermediate principal axes of the linear tidal field, 
which disseminates a prospect of using them as a probe of early universe physics. This roseate prospect, however, is contingent upon the 
key assumption that the observable stellar spins of the present galaxies measured at inner radii have the same alignment tendency toward the initial tidal field 
as their dark matter counterparts measured at virial limits. We test this assumption directly against a high-resolution hydrodynamical simulation by tracing back 
the galaxy component particles back to the protogalactic stage. It is discovered that the galaxy stellar spins at $z=0$ have strong but {\it reoriented} memory 
for the early universe, exhibiting a significant signal of cross-correlation with the {\it major} principal axes of the initial tidal field at $z=127$. 
An analytic single-parameter model for this reorientation of the present galaxy stellar spins relative to the initial tidal field is devised and shown to be in good accord 
with the numerical results. 
\end{abstract}
\section{Introduction}\label{sec:intro}

The {\it directions of galaxy angular momentum vectors}, which will be referred to as the {\it galaxy spins} throughout this paper, 
have been regarded as one of those very few near-field observables that are analytically tractable within the linear perturbation theory \citep{whi84,CT96}, 
which generically predicts preferential alignments of the proto-galaxy spins with the intermediate principal axes of the initial tidal field \citep{LP00,LP01,por-etal02a,por-etal02b}.  
This version of the linear perturbation theory adapted for the origin of the galaxy angular momentum is often dubbed the linear tidal torque theory (LTTT).  
The initially acquired alignment tendency predicted by the LTTT is believed to be well retained during the subsequent nonlinear evolution 
\citep[e.g.,][]{LE07,zha-etal15,lee-etal18,mot-etal21}, even though the galaxy spins swing and flip with respect to the cosmic web through 
hierarchical merging process \citep[e.g.,][]{ara-etal07,hah-etal07,paz-etal08,cod-etal12,tro-etal13,lib-etal13,dub-etal14,cod-etal15,WK17,cod-etal18,gan-etal18,
wan-etal18,kra-etal20,lee-etal20,LL20}. 

Several $N$-body simulations have indeed confirmed the existence of the alignments between the spins of galactic halos and the intermediate 
principal axes of the local tidal field even in the nonlinear regime ($z\le 2$), although the alignment tendency gradually becomes weaker at lower redshifts 
and on the lower mass scales \citep[e.g.,][]{for-etal14,che-etal16,lee-etal20,lee-etal21}. 
These numerical evidences inspired several authors to investigate a possibility of probing the physics of early universe with the galaxy spins. 
For instance, \citet{yu-etal20} showed that the degree of chiral asymmetry could in principle be constrained by measuring the 
present galaxy spins. \citet{mot-etal21} suggested that the galaxy spins be a useful diagnostics of cosmic inflation and primordial 
gravitational waves. \citet{LL20} found that the threshold mass at which the alignments between the galaxy spins and local 
tidal field become negligibly weak has a potential to distinguish between viable dark energy candidates. 
The success of this new probe, however, depends entirely on how strongly the present galaxy {\it stellar} spins are cross-correlated with the initial tidal  
field, which is not necessarily guaranteed by their correlations with the local tidal fields, i.e., surrounding cosmic web. 
 
\citet{mot-etal21} reported a $2.7\sigma$ signal of the alignments between the present galaxy stellar spins and reconstructed initial galaxy dark 
matter (DM) spins from the observational data. 
The former was determined from the Sloan Digital Sky Survey (SDSS) galaxies whose morphologies had been classified by the Galaxy Zoo project \citep{zoo}, 
while the latter was evaluated according to the LTTT with the initial density and tidal fields reconstructed by the {\tt ELUCID} algorithm \citep{elucid}.
Although this observational signal supports the conservation of the galaxy spins, it still cannot be directly translated into the existence 
of the alignments between the present galaxy stellar spins and the intermediate principal directions of the initial tidal field.

Furthermore, a recent numerical experiment based on a high-resolution hydrodynamical simulation found that the galaxy spins exhibit a radius dependent transition with 
respect to the local tidal field \citep{ML23}.  Unlike the galaxy {\it virial} spins measured at the virial radii, $\rv$, the galaxy inner spins measured at inner radii $r<\rv$ 
were found to be aligned with the {\it major} principal axes of the local tidal field, depending on the scale. 
In true observations, the galaxy stellar spins can be measured only at $r\lesssim 2\rh$ \citep[e.g.,][]{rom-etal03,coc-etal09} with the stellar half-mass radius $\rh$, much more 
inner regions than the virial limits. 
Meanwhile, what is predicted to be aligned with the intermediate principal directions of the initial tidal field is the galaxy virial spins \citep{LP00,LP01}.  

The detection of the radius dependent spin transition phenomena gives a warning to any parallel comparison between the observable galaxy stellar spins and 
the predictions of LTTT. It is an essential prerequisite to explore if and how the observable stellar spins of the present galaxies measured at $\lesssim 2\rh$ are truly 
aligned with the principal axes of the initial tidal field smoothed on the galactic scale $\sim 0.5\dunit$ \citep{bbks}, before using the galaxy spins as a new window 
on the early universe. In this Paper, we set out to fulfill this core mission.

\section{Alignments between Present Galaxy Spins and Initial Tidal Fields}

\subsection{Numerical Analysis}\label{sec:num}

We make use of the subhalos resolved via the SUBFIND algorithm \citep{subfind} in the TNG300-1 hydrodynamical 
simulation of the IllustrisTNG project \citep{tngintro1,tngintro2,tngintro3,tngintro4,tngintro5,illustris19}, referring to them as the galaxies. 
The simulation was performed on a periodic cubic box of a side length $205\dunit$ with a total of 
$2\times 2500^{3}$ elements for the Planck $\lcdm$ (cosmological constant $\Lambda$+cold dark matter) universe \citep{planck16}. 
A half of the elements behaves as collisionless DM particles that only gravitationally interact, while the other half is the collisional baryons that can cool down to form stars. 
The average mass resolutions of the DM and baryonic elements attain $5.9\times 10^{7}\,M_{\odot}$ and $1.1\times 10^{7}\,M_{\odot}$, respectively.
See the IllustrisTNG webpage for a full information on the codes and hydrodynamical physics implemented into the simulation. 

Applying the cloud-in-cell method to the spatial distributions of all DM and baryon elements at $z=127$ where the simulation starts, 
we first construct a real-space initial density field, $\delta({\bf q})$, on the $512^{3}$ grids and then Fourier-transform 
it into $\delta({\bf k})$, with wave vector ${\bf k}=k(\hat{k}_{l})$.  The initial tidal field, ${\bf T}\equiv (T_{ln})$, convolved with a Gaussian kernel of scale radius $\rf$, 
is then determined via an inverse Fourier transformation of $\hat{k}_{l}\hat{k}_{n}\exp(-k^{2}R^{2}_{f}/2)\delta({\bf k})$ as in \citet{ML23}. 
Throughout this paper, we will consider two different kernel scales, $\rf/(\dunit)=0.5$ and $1$.

For the current analysis, a total of $44956$ galaxies are selected at $z=0$, which have total masses $M_{g}$ in a range of $11.8\le\log\mg\le 12.6$ with 
$\mg\equiv M_{g}/(\munit)$ and contain $300$ or more stellar particles ($n_{s}\ge 300$) within $2\rh$ \citep{bet-etal07}. The present stellar spin of each selected galaxy is 
determined as $\js\equiv {\bf J}_{\star}(z=0)/\vert{\bf J}_{\star}(z=0)\vert$ with 
${\bf J}_{\star}(z=0)\equiv\sum_{\gamma=1}^{n_{s}}{m_{\star\gamma}}\,({\bf x}_{c}-{\bf x}_{\gamma})\times({\bf v}_{c}-{\bf v}_{\gamma})$, where 
$m_{\star\gamma}$, ${\bf x}_{\gamma}$ and ${\bf v}_{\gamma}$  denote the mass, comoving position and peculiar velocity of the $\gamma$th stellar particle, 
respectively, while ${\bf x}_{c}$ and ${\bf v}_{c}$ are the comoving position and peculiar velocity of the potential minimum (coinciding well with the galaxy 
center of mass), respectively. 

\begin{figure}[ht]
\centering
\includegraphics[height=18cm,width=14cm]{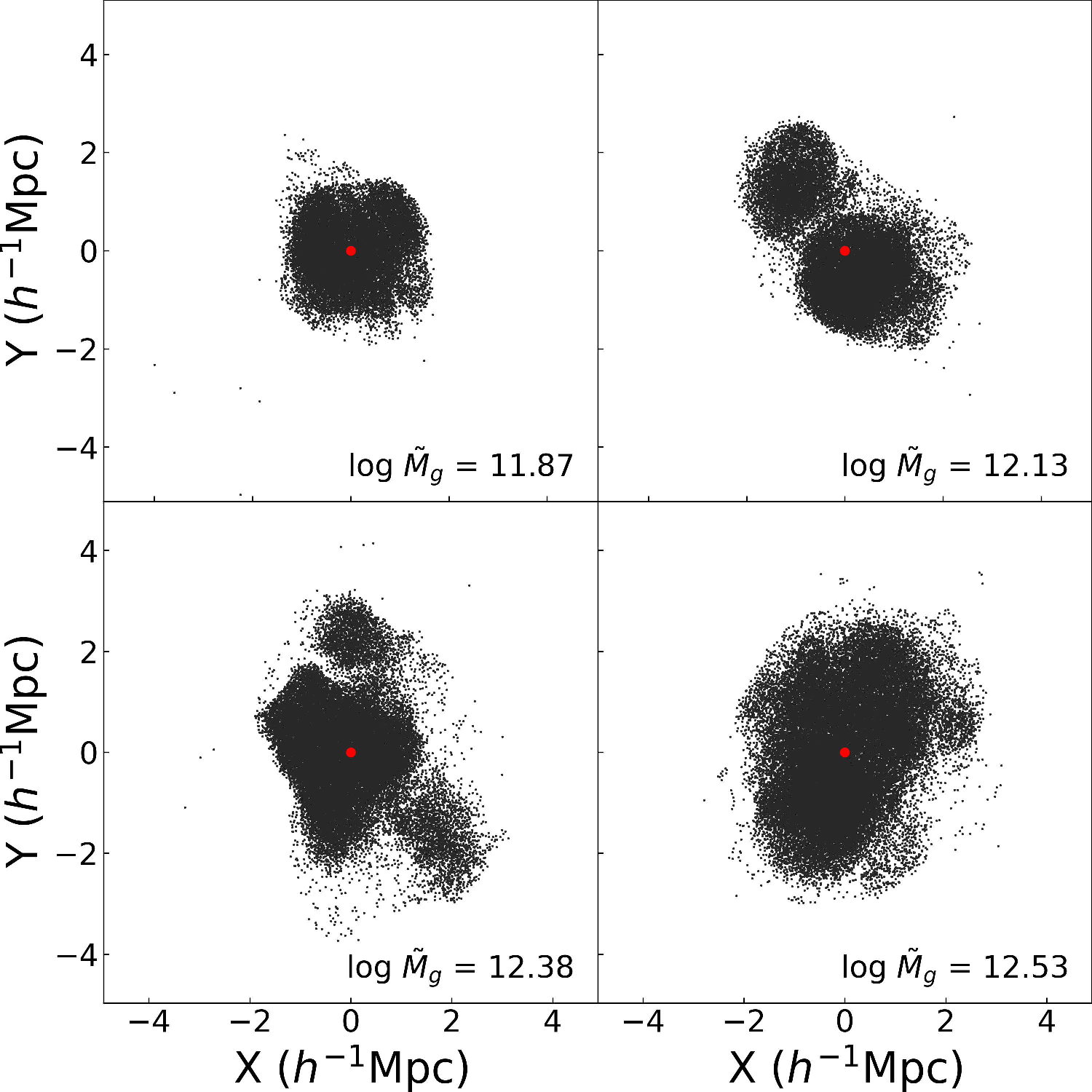}
\caption{Two dimensional projected images of randomly selected four proto-galaxies in the $\hat{\bf x}$-$\hat{\bf y}$ plane at $z=127$.} 
\label{fig:image}
\end{figure}

Tracing all constituent DM particles of each selected galaxy enclosed within $\rv$ back to $z=127$, we find its proto-galactic sites and locate its initial center of mass, 
${\bf q}_{c}$.  Figure \ref{fig:image} shows the two dimensional projected images of randomly selected proto-galaxies with their center of masses marked by red dots 
in four different mass sections. As can be seen,  the spatial extents of the lowest-mass proto-galaxies are larger than the smoothing scales, $\rf/(\dunit)\le 1$, which is 
important to ascertain since the initial tidal field smoothed on the scales larger than the spatial extents of the proto-galactic sites could produce overdamped alignment signals. 
The initial tidal field, ${\bf T}({\bf q}_{c})$, is put through a similarity transformation to yield its three 
orthonormal eigenvectors, $\{\tm({\bf q}_{c}),\ti({\bf q}_{c}),\tn({\bf q}_{c})\}$, parallel to its major, intermediate and minor principal axes, respectively.  
The alignments of the present galaxy stellar spins and the principal axes of the initial tidal field as a function of $\log\mg$ are then evaluated as 
$\{\langle\vert\js\cdot\hat{\bf t}_{l}\vert\rangle\}_{l=1}^{3}$, where the ensemble averages are taken over the galaxies whose masses fall in a given 
differential interval of $[\log\mg, \log\mg+d\log\mg]$. 

\begin{figure}[ht]
\centering
\includegraphics[height=18cm,width=14cm]{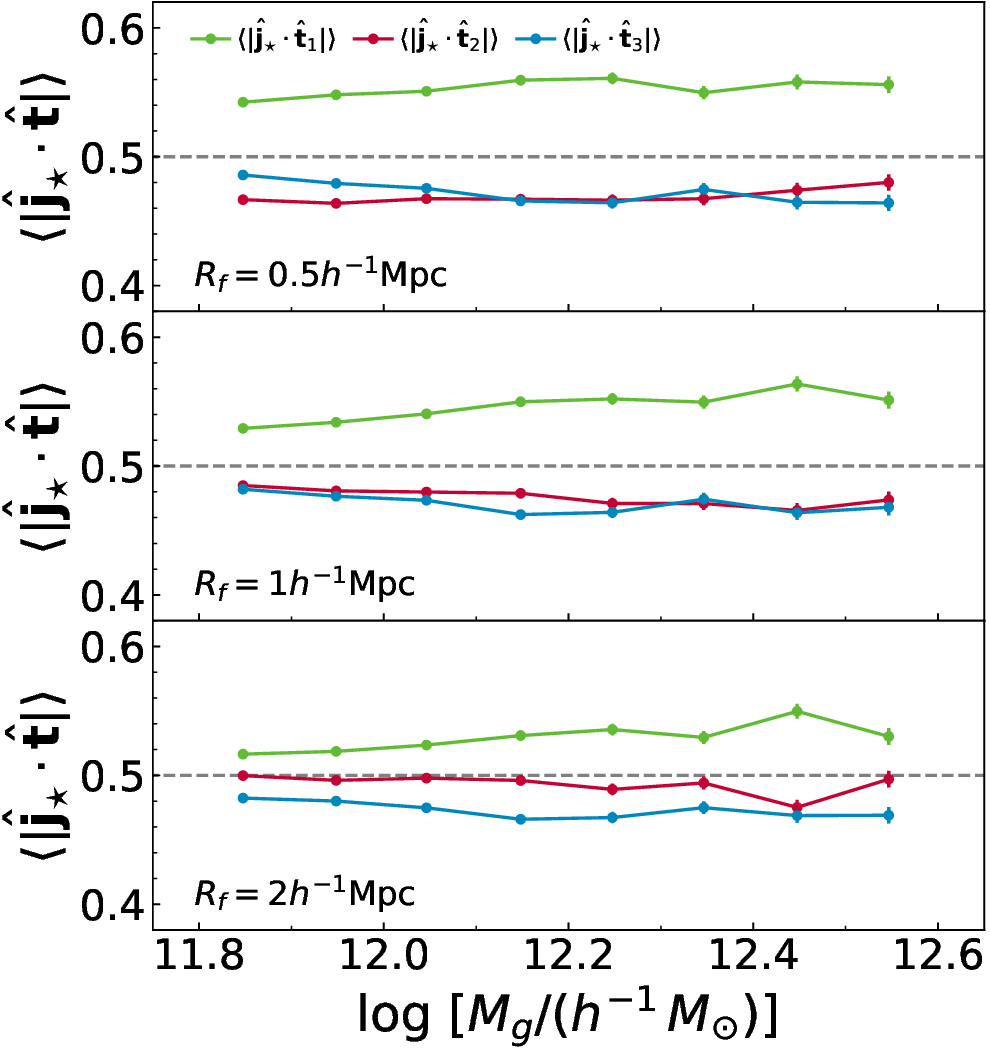}
\caption{Alignments of the present galaxy stellar spins at $z=0$ and three principal axes of the initial tidal field at $z=127$ as a 
function of the logarithms of the total galaxy mass for three different cases of the smoothing scale. In line with observations, 
the present galaxy stellar spins are measured at $2\rh$.} 
\label{fig:js0}
\end{figure}

In Figure \ref{fig:js0} we show $\{\langle\vert\js\cdot\tm\vert\rangle$ (green line), $\langle\vert\js\cdot\ti\vert\rangle$ (red line), 
$\langle\vert\js\cdot\tn\vert\rangle\}$ (blue line) versus $\log\mg$ for three different cases of $\rf/(\dunit) = 0.5$, $1$ and $2$ in the top, middle 
and bottom panels, respectively.  The associated errors are computed as one standard deviation in the ensemble averages. 
As can be seen, the present galaxy stellar spins are preferentially aligned with the major (rather than intermediate) principal axes with the initial tidal field for 
all of the three cases of $\rf$, unlike the expectation based on the LTTT and angular momentum conservation \citep{LP00,LP01}. 
Note that the $\js$--$\tm$ alignment mildly dwindles as $\log\mg$ decreases and as $\rf$ increases. 

\begin{figure}[ht]
\centering
\includegraphics[height=18cm,width=14cm]{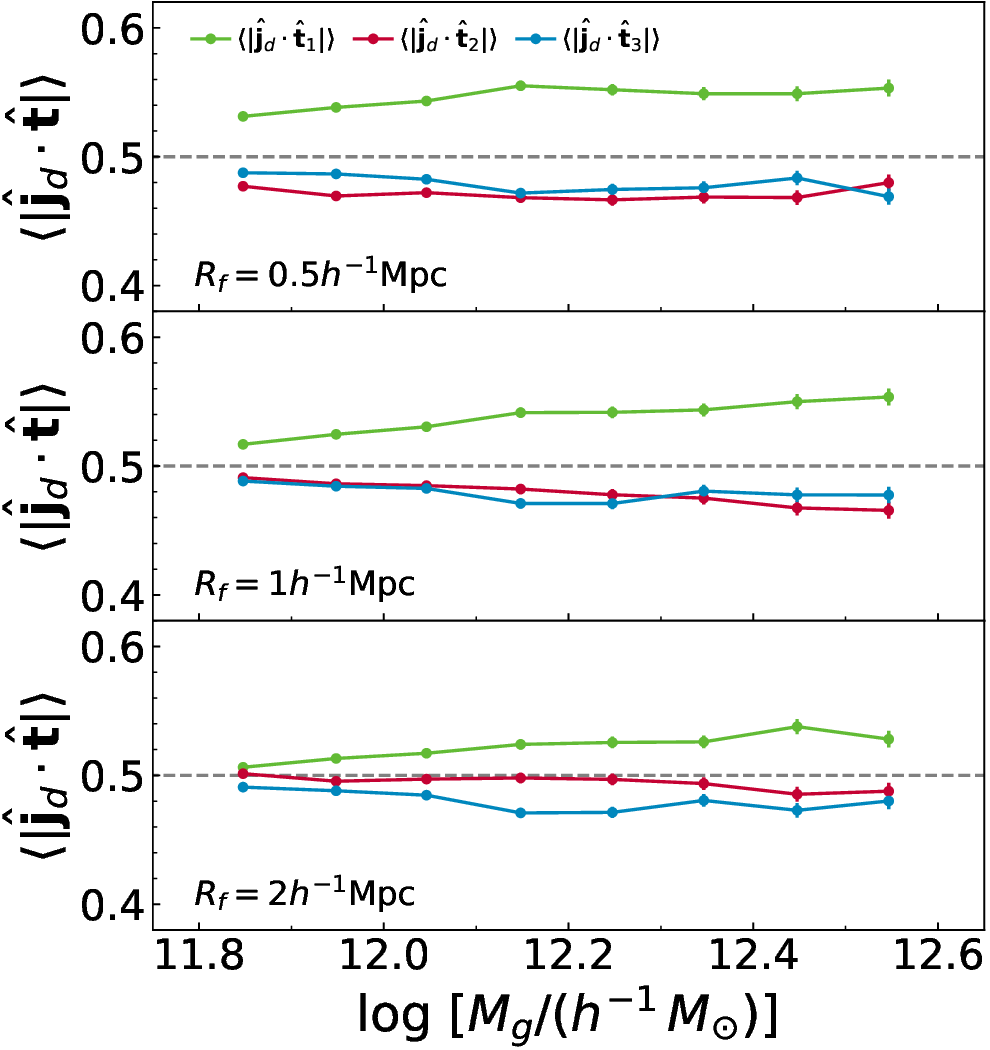}
\caption{Same as Figure \ref{fig:js0} but with the present galaxy DM inner spins measured at $2\rh$.}
\label{fig:jd0}
\end{figure}

To see whether or not this somewhat unexpected $\js$--$\tm$ alignment is caused by some baryonic effects in the subsequent evolution, we measure the 
present galaxy DM spins, $\jd$, from the DM particles enclosed within $2\rh$ and repeat the same analysis, the results of which are shown 
in Figure \ref{fig:jd0}. As can be seen, little difference exists between $\jd$ and $\js$ in the tendencies and strengths of their alignments with 
$\{\hat{\bf t}_{l}({\bf q}_{c})\}_{l=1}^{3}$, which clearly rules out any baryon effects as a possible origin of the $\js$--$\tm$ alignment. 

\begin{figure}[ht]
\centering
\includegraphics[height=18cm,width=14cm]{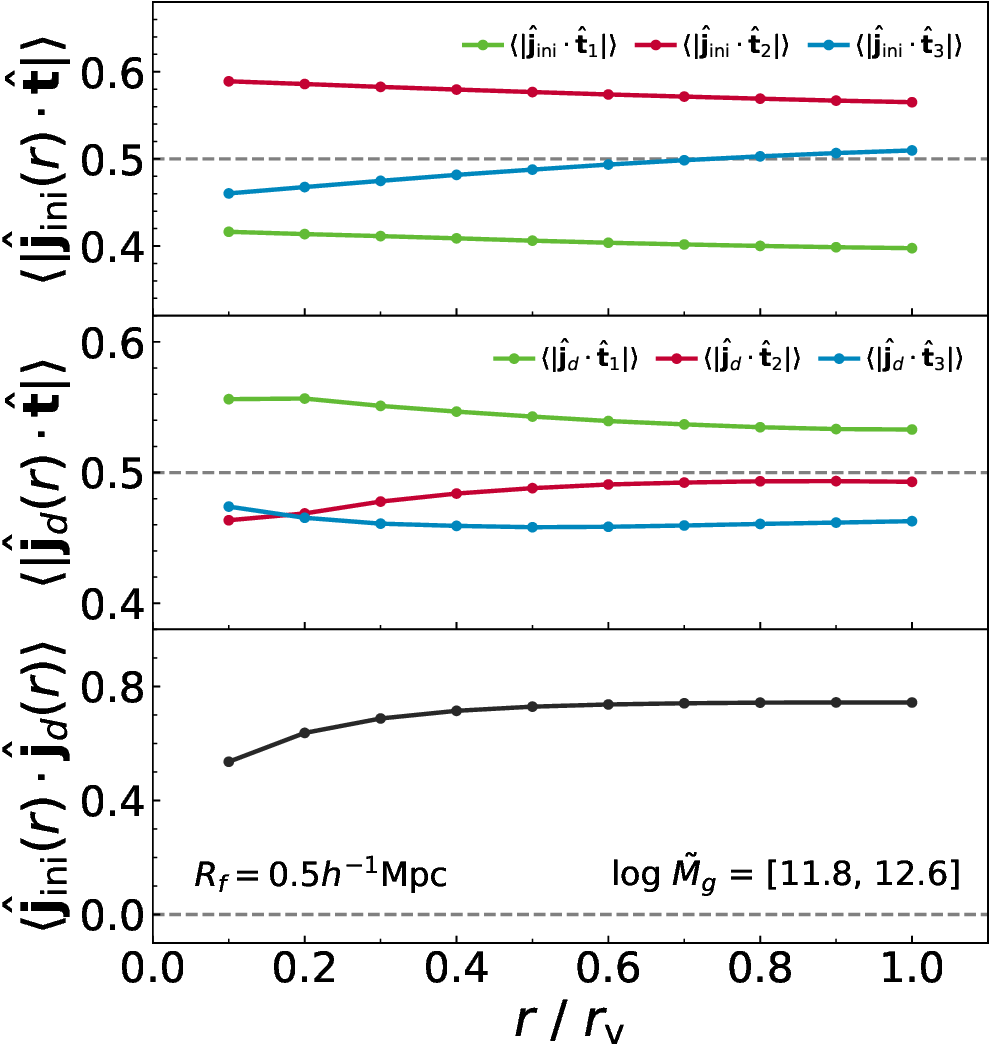}
\caption{(Top panel): alignments between the initial galaxy spins measured at inner radii $r\le \rv$ and three principal axes of the initial tidal field at the 
same epoch as a function of the inner to virial radius ratio for the case of $\rf=0.5/(\dunit)$. (Middle panel): same as the top panel but with the present galaxy 
spins at $z=0$. (Bottom panel): alignments between the initial and present galaxy spins.}
\label{fig:jd127}
\end{figure}

To test the LTTT whose archetypal prediction was the alignments between the {\it initial} galaxy spins and the intermediate principal axes of the initial tidal field,
we measure the initial galaxy DM spins, $\jz\equiv \hat{\bf j}(z=127)$, from the initial positions and peculiar velocities of the DM particles 
that end up being within $r\le\rv$ from the present galaxy centers at $z=0$, and investigate their alignments with the principal axes of the initial tidal field, 
$\{\langle\vert\jz\cdot\hat{\bf t}_{l}\vert\rangle\}_{l=1}^{3}$, as a function of $r/\rv$. 
The top panel of Figure  \ref{fig:jd127} plots $\{\langle\vert\jz(r)\cdot\hat{\bf t}_{l}\vert\rangle\}_{l=1}^{3}$ versus $r/\rv$ for the case of 
$\rf/(\dunit)=0.5$, demonstrating that the initial galaxy spins are indeed well aligned with the intermediate principal axes of the initial tidal field in the whole 
range of $0\le r/\rv\le 1$. This result confirms that the LTTT predictions hold true not only for the virial spins but also for the inner spins as far as they 
are measured at the initial epochs. 
For comparison, the radius dependence of the alignment tendency of the present galaxy DM spins, $\jd (r)$, with the initial tidal field is also investigated and 
shown in the middle panel of Figure \ref{fig:jd127}. Note that $\jd(r)$ are always aligned with $\tm$ not only for the case of $r<\rv $ but also for the case $r=\rv$. 

Questioning if this obvious difference between the present and initial galaxy spins in their alignment tendencies toward the initial tidal field indicates a complete 
breakdown  of angular momentum conservation, we also compute the alignments between the present and initial galaxy DM spins, $\langle\jd(z)\cdot\jz(r)\rangle$ 
as a function of $r/\rv$, which are shown in the bottom panel of Figure  \ref{fig:jd127}. As can be seen, the present galaxy spins are indeed strongly aligned with the 
initial galaxy spins, confirming that the galaxy angular momentum are quite well conserved even during the nonlinear evolution. 

A central implication of the results shown in Figs. \ref{fig:js0}--\ref{fig:jd127} is the following. Despite that the LTTT is indeed quite successful in describing the 
correlations between the galaxy spins and the tidal field measured at the same initial epochs and that the present galaxy spins are well aligned with the initial 
counterparts,  the validity of the LTTT cannot be extrapolated to predict the cross-correlations between the present galaxy inner spins (or stellar spins) and the initial 
tidal field since the present galaxy inner spins are reoriented from their initial versions. The more inner radii the present spins are measured at, the more strongly their 
alignment tendency deviates from the prediction of the LTTT. 

\subsection{Analytical Modeling}\label{sec:analytic}

For a better understanding of the numerical results presented in Section \ref{sec:num},  we construct an analytic model for the probability density function, 
$p(\vert\js\cdot\tm\vert)$. As in \citet{LP01}, we assume that the conditional joint probability density function of three coordinates of the angular momentum vectors 
of the galaxy stellar components given a tidal field, $p(J_{\star 1}, J_{\star 2}, J_{\star 3}\vert \tilde{\bf T})$, can be well approximated by  the following multivariate 
Gaussian distribution \citep[see also][]{lee19}
\begin{equation}
\label{eqn:jpro}
p(J_{\star 1},\ J_{\star 2},\ J_{\star 3}\vert\tilde{\bf T}) = \frac{1}{\sqrt{(2\pi)^3 {\rm det}({\bf \Sigma})}}
\exp\left[-\frac{1}{2}\left({\bf J}^{t}_{\star}\cdot{\bf \Sigma}^{-1}\cdot{\bf J}_{\star}\right)\right]\, , 
\end{equation}
where $\tilde{\bf T}$ is the traceless version of ${\bf T}$, and ${\bf \Sigma}=({\Sigma}_{ln})$ is the covariance matrix whose components are given as
${\Sigma}_{ln} \equiv \langle{J}_{\star l}{J}_{\star n}\vert\tilde{\bf T}\rangle$. 

Marginalize $p(J_{\star 1}, J_{\star 2}, J_{\star 3}\vert\tilde{\bf T})$ over the magnitude of ${\bf J}_{\star}$, we compute the conditional probability density 
function of the galaxy stellar spins as 
\begin{eqnarray}
\label{eqn:hjpro}
p(\hat{j}_{\star 1},\ \hat{j}_{\star 2},\ \hat{j}_{\star 3}\vert\hat{\bf T})&=&\int\,p({\bf J}_{\star}\vert\tilde{\bf T}) J^{2}_{\star}\,dJ_{\star} \nonumber \\
&=&\frac{1}{4\pi\left[{\rm det}(\hat{\bf \Sigma})\right]^{1/2}}\left(\hat{\bf j}_{\star}\cdot\hat{\bf \Sigma}^{-1}\cdot\hat{\bf j}_{\star}\right)^{-3/2}\, ,
\end{eqnarray}
where $\hat{\bf \Sigma}=(\hat{\Sigma}_{ln})$ is the unit covariance matrix, defined as $\hat{\Sigma}_{ln}=\langle\hat{j}_{\star l}\hat{j}_{\star n}\vert\hat{\bf T}\rangle$ 
with the unit traceless tidal field $\hat{\bf T}\equiv \tilde{\bf T}/\vert\tilde{\bf T}\vert$ \citep[see,][]{LP00,LP01,lee19}. 

Here, we propose an empirical, parameterized model for $\hat{\bf \Sigma}$: 
\begin{equation}
\label{eqn:cov}
\langle\hat{\bf j}_{\star l}\hat{\bf j}_{\star n}\vert\hat{\bf T}\rangle \approx \frac{1+\kappa}{3}\delta_{ln}-\kappa\hat{R}_{lu}\hat{T}_{un}\, , 
\end{equation}
where $\delta_{ln}$ is the Kronecker-delta symbol, the parameter $\kappa$ is a measure of the strength of the cross-correlation 
between $\js$ and $\hat{\bf T}$, and $\hat{\bf R}\equiv (\hat{R}_{ln})$ is a rotation matrix with $\vert\hat{\bf R}\vert = 1$. 
In the original work of \citet{LP01} who constructed an analytical model for the tidally induced spin alignments of the galactic halos, the unit covariance matrix, 
$\hat{\bf \Sigma}$, was modeled as $\langle\hat{\bf j}_{l}\hat{\bf j}_{n}\vert\hat{\bf T}\rangle \approx (1+\kappa)\delta_{ln}/3 - \kappa\hat{T}_{lu}\hat{T}_{un}$, which 
naturally leads to the prediction of $\jd$-$\ti$ alignments. This original model, however, turned out to be valid only provided that the tidal fields and galaxy virial 
spins are measured at the same epochs.

To accommodate the cross-correlations between the present galaxy inner spins and the initial tidal fields, we make an ad-hoc modification of their model  into 
Equation (\ref{eqn:cov}) by replacing $\hat{T}_{lu}\hat{T}_{un}$ by $\hat{R}_{lu}\hat{T}_{un}$. Basically, this rotation matrix,  $\hat{\bf R}$, quantifies how the present 
galaxy spins are reoriented from the initial galaxy spins in the principal frame of the initial tidal field. 
Given the $\js$-$\tm$ alignments, we model $(\hat{R}_{ln})$ to represent a maximum $45^{\circ}$ rotation upon the $\tm$-axis in the plane spanned by $\ti$ and $\tn$, 
i.e, $\hat{R}_{11}=1$, $\hat{R}_{22}=\hat{R}_{23}=\hat{R}_{33}=-\hat{R}_{32}=1/\sqrt{2}$, and $\hat{R}_{12}=\hat{R}_{13}=\hat{R}_{21}=\hat{R}_{31}=0$.

Inserting Equation (\ref{eqn:cov}) into Equation (\ref{eqn:hjpro}), we now express the conditional probability density\footnote{In the work of \citet{lee19}, 
the analytic expression for $p(\hat{j}_{1},\ \hat{j}_{2},\ \hat{j}_{3}\vert\hat{\bf T})$ has a typo error, $\vert\hat{\bf j}\cdot\hat{\bf t}\vert$. It is correctly fixed into 
$\vert\hat{\bf j}\cdot\hat{\bf t}\vert^{2}$ here.} of the galaxy stellar spins in the principal frame of $\hat{\bf T}$
\begin{eqnarray}
\label{eqn:hjpro_axis}
p(\hat{j}_{\star 1},\ \hat{j}_{\star 2},\ \hat{j}_{\star 3}\vert\hat{\bf T}) 
&=& \frac{1}{2\pi}\left[\prod_{n=1}^{3}\left(1+\kappa-3\kappa\hat{\varrho}_{n}\right)\right]^{-\frac{1}{2}}
\left(\sum_{l=1}^{3}\frac{\vert\js\cdot\hat{\bf t}_{l}\vert^{2}}{1+\kappa-3\kappa\hat{\varrho}_{l}} \right)^{-\frac{3}{2}}\, ,
\end{eqnarray}
where $\{\hat{\varrho}\}_{l=1}^{3}$ are the eigenvalues of the symmetric diagonal maxtrix, 
$\hat{\bf R}_{lu}\hat{\bf T}_{un}$. Recalling that the eigenvalues of $\hat{\bf T}$ has the ensemble averages of 
$\langle\hat{\lambda}_{1}\rangle\approx 1/\sqrt{2}$, $\langle\hat{\lambda}_{2}\rangle\approx 0$, and 
$\langle\hat{\lambda}_{3}\rangle \approx -1/\sqrt{2}$ \citep{LP01}, we expect 
$\langle\hat{\varrho}_{1}\rangle\approx 1/\sqrt{2}$, $\langle\hat{\varrho}_{2}\rangle\approx -1/2$, and 
$\langle\hat{\varrho}_{3}\rangle \approx -1/2$.

Let $\theta_{1}$ and $\phi_{1}$ denote the spherical polar and azimuthal angles of $\js$, respectively, in the Cartesian frame whose $\hat{\bf x}$, $\hat{\bf y}$, 
and $\hat{\bf z}$ axes are chosen to be parallel to $\ti$, $\tn$ and $\tm$, respectively. 
That is, $\theta_{1}\equiv \vert\js\cdot\tm\vert$, while $\phi_{1}$ is the angle of $\js$ projected onto the 
plane spanned by $\ti$ and $\tn$.  We now express $p(\js\vert\hat{\bf T})$ in terms of $\cos\theta_{1}$ and $\phi_{1}$ as
\begin{eqnarray}
\label{eqn:hjpro_prin}
p(\cos\theta_{1},\phi_{1}) 
&=& \frac{1}{2\pi}\left[\prod_{n=1}^{3}\left(1+\kappa-3\kappa\hat{\varrho}_{n}\right)\right]^{-\frac{1}{2}}\times \nonumber \\
&&\left[\frac{\cos^{2}\theta_{1}}{1+\kappa-3\kappa\hat{\varrho}_{1}} + \frac{\left(1-\cos^{2}\theta_{1}\right)\cos^{2}\phi_{1}}{1+\kappa-3\kappa\hat{\varrho}_{2}}  + 
\frac{\left(1-\cos^{2}\theta_{1}\right)\sin^{2}\phi_{1}}{1+\kappa-3\kappa\hat{\varrho}_{3}} \right]^{-\frac{3}{2}}\, ,
\end{eqnarray}

Putting these ensemble average values into Equation (\ref{eqn:hjpro_prin}) and marginalizing it over $\phi_{1}$, we have  
\begin{eqnarray}
\label{eqn:pro0}
p(\cos\theta_{1})&=&\int_{0}^{2\pi}p(\js\vert\hat{\bf T})\,d\phi \nonumber \\
\label{eqn:pro1}
&=& A\left[\frac{\cos^{2}\theta_{1}}{1+(1-3/\sqrt{2})\kappa} + \frac{1-\cos^{2}\theta_{1}}{1+5\kappa/2}\right]^{-\frac{3}{2}}\, ,
\end{eqnarray}
where $A\equiv [(1+(1-3/\sqrt{2})\kappa)\left(1+5\kappa/2\right)^{2}]^{-\frac{1}{2}}$ \citep{lee19}. 

The other two probability densities are also derived as 
\begin{eqnarray}
\label{eqn:pro2}
p(\cos\theta_{2})&=&\frac{A}{2\pi}\!\int_{0}^{2\pi}d\phi_{2}\Bigg{\{}\frac{\cos^{2}\theta_{2}+\left(1-\cos^{2}\theta_{2}\right)\cos^{2}\phi_{2}}{1+5\kappa/2} + 
\frac{\left(1-\cos^{2}\theta_{2}\right)\sin^{2}\phi_{2}}{1+(1-3/\sqrt{2})\kappa}\Bigg{\}}^{-\frac{3}{2}}, \\
\label{eqn:pro3}
p(\cos\theta_{3})&=&\frac{A}{2\pi}\!\int_{0}^{2\pi}d\phi_{3}\Bigg{\{}\frac{\cos^{2}\theta_{3}+\left(1-\cos^{2}\theta_{3}\right)\sin^{2}\phi_{3}}{1+5\kappa/2} + 
\frac{\left(1-\cos^{2}\theta_{3}\right)\cos^{2}\phi_{3}}{1+(1-3/\sqrt{2})\kappa}\Bigg{\}}^{-\frac{3}{2}}\, ,
\end{eqnarray}
with $\cos\theta_{2}\equiv\vert\js\cdot\ti\vert$, $\cos\theta_{3}\equiv \vert\js\cdot\tn\vert$, and $\{\phi_{2},\phi_{3}\}$ are the azimuthal angles of $\js$ projected onto 
the $\hat{\bf t}_{3}$-$\hat{\bf t}_{1}$ and $\hat{\bf t}_{1}$-$\hat{\bf t}_{2}$ planes, respectively. 

\begin{figure}[ht]
\centering
\includegraphics[height=18cm,width=14cm]{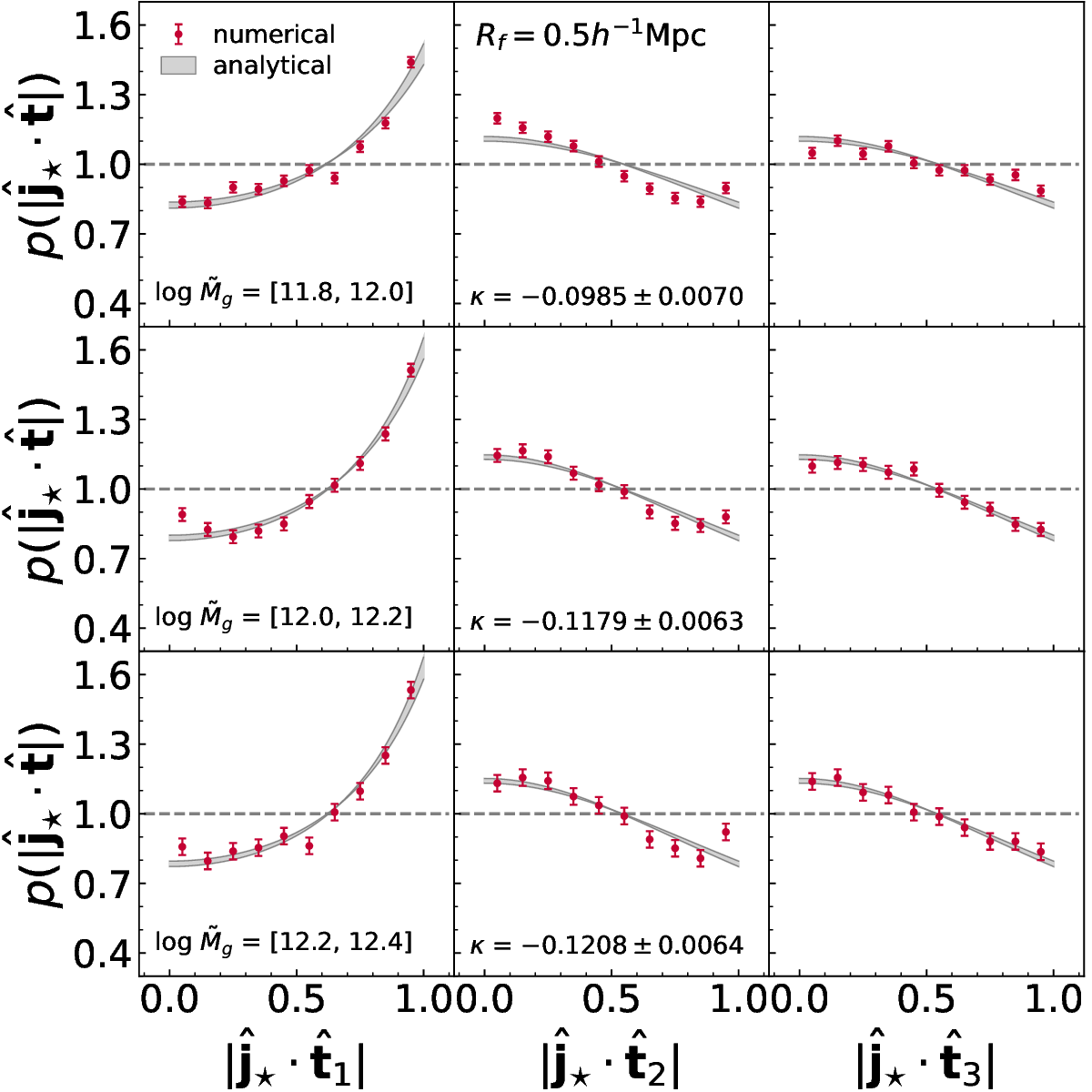}
\caption{Analytic formula (gray areas) for the probability densities of the alignments between the present galaxy stellar spins and the 
three principal axes of the initial tidal field in three different mass ranges, compared with the numerical results (red filled circles) for 
the case of $\rf/(\dunit)=0.5$.}
\label{fig:pro0.5}
\end{figure}

\begin{figure}[ht]
\centering
\includegraphics[height=18cm,width=14cm]{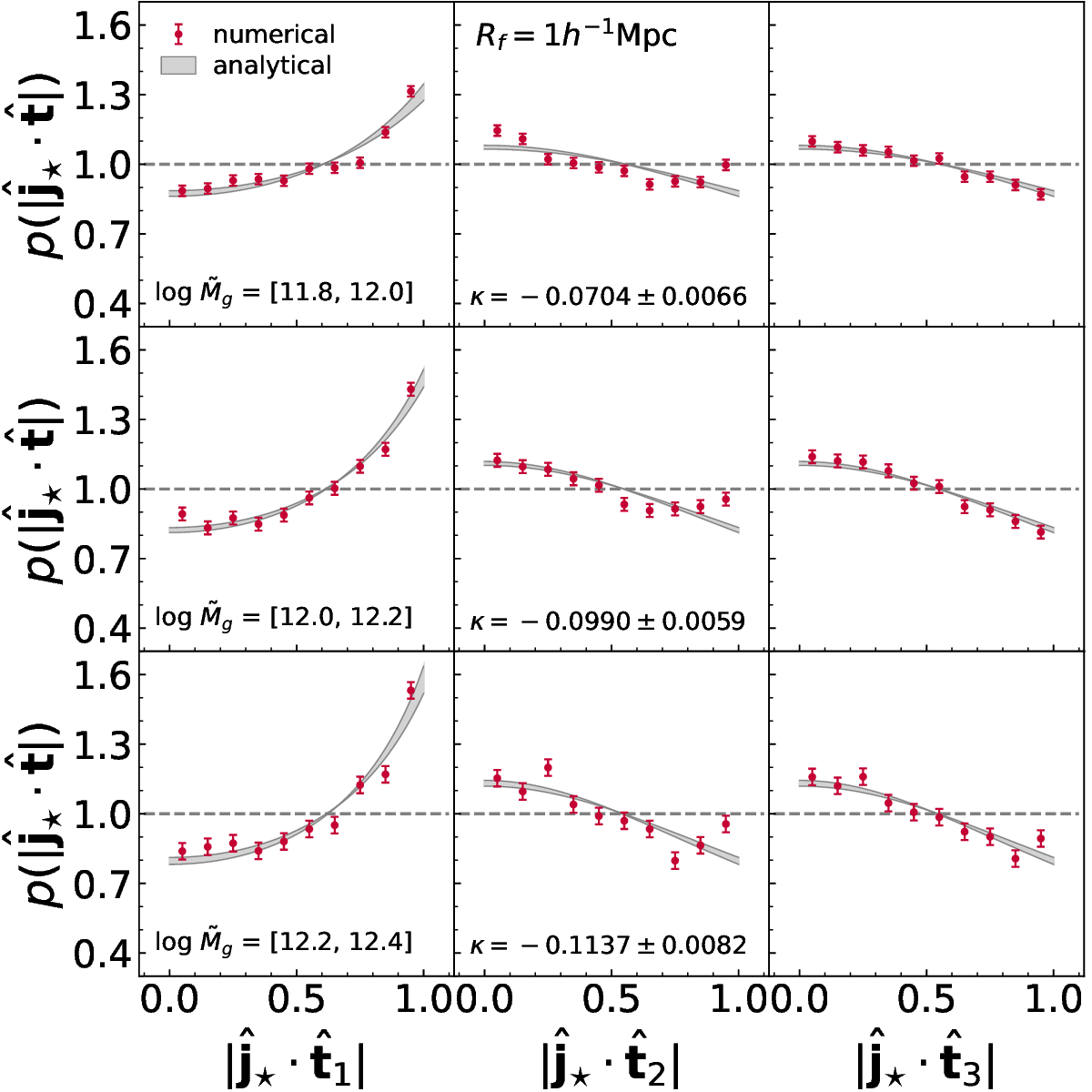}
\caption{Same as Figure \ref{fig:pro0.5} but for the case of $\rf/(\dunit)=1$.}
\label{fig:pro1}
\end{figure}

In Figures \ref{fig:pro0.5}-\ref{fig:pro1} we demonstrate how well Equations (\ref{eqn:pro1})--(\ref{eqn:pro3}) (gray areas) agree with the numerically obtained 
$\{p(\vert\js\cdot\hat{\bf t}_{l}\vert)\}_{l=1}^{3}$ (red filled circles) with Poisson errors in three different $\log\mg$ ranges 
for the cases of $\rf/(\dunit)=0.5$ and $1$, respectively.  The one standard deviation confidence interval of $\kappa$ is determined by fitting the numerical result of 
$p(\vert\js\cdot\tm\vert)$ to Equations (\ref{eqn:pro1})--(\ref{eqn:pro3}) via the $\chi^{2}$-minimization for each case. 
The good agreements between the analytical and numerical results are found to be robust against the variations of $\mg$ and $\rf$, proving the practical usefulness 
of Equation (\ref{eqn:cov}).  Given that a more directly measurable quantity in practice is not the total mass of a galaxy but its stellar mass, we provide information on the 
mean stellar masses corresponding to the three $\mg$ ranges: 
$\langle\log M_{\star}/(\munit)\rangle =10.06$, $10.27$, and $10.43$, respectively.

\section{Summary and Conclusion}\label{sec:sum}

Analyzing the subhalo catalogs in the galactic mass range from the TNG300-1 hydrodynamic simulation \citep{tngintro1,tngintro2,tngintro3,tngintro4,tngintro5,illustris19}, 
we have discovered that the present galaxy stellar spins ($\js$) measured at the inner radii $2\rh$ are preferentially aligned with the directions 
parallel to the {\it major} principal axes ($\tm$) and perpendicular to the planes spanned by the intermediate and minor principal axes ($\ti$ and $\tn$, respectively) of the 
initial tidal field smoothed on the scales of $\rf/(\dunit)=0.5$ and $1$. 
The baryonic effects have been ruled out as a possible cause of the detected $\js$-$\tm$ alignments since the DM spins measured at $2\rh$ exhibit the same alignment 
tendency.

We put forth the following scenario to understand the observed $\js$-$\tm$ alignments. The initial tidal field exerts {\it torque wrench} on the proto-galaxies, 
which cause them to become compressed as well as to rotate.  Before the turn-around moments, however, the proto-galaxies experience only rotations, 
developing the $\jz$-$\ti$ alignments as predicted by \citet{LP00}, since the tension produced by the expanding space-time completely cancels the compression. 
However, in the post turn-around era when the tension effect diminishes, the net compression effect begins to predominate especially 
in the galaxy inner regions. In consequence, the galaxy stellar spins become reoriented from the initial tendency, developing a preferential alignment toward $\tm$.
Under this scenario, we have come up with an analytic single -parameter formula for the probability density functions, $\{p(\vert\js\cdot\hat{\bf t}_{l}\vert)\}_{l=1}^{3}$, 
with an ad-hoc assumption that the reorientation of the present galaxy stellar alignments can be described by a $45^{\circ}$ rotation of the principal frame of the initial tidal 
field upon the $\tm$-axis, and found that the formula with the best-fit values of the parameter agree quite well with the numerically determined  
$\{p(\vert\js\cdot\hat{\bf t}_{l}\vert)\}_{l=1}^{3}$ for both of the cases of $\rf$.

Despite that our result challenges the standard picture based on the LTTT combined with the conservation of galaxy spins which predicts the 
$\js$--$\ti$ alignments \citep[e.g.,][]{LP00,LP01,yu-etal20,mot-etal21}, it does not controvert but rather corroborate the power of the present galaxy stellar spins 
as a cosmological probe owing to their even stronger memory for the early universe than expected by the LTTT. To properly probe the early universe physics 
with observable galaxy stellar spins, however, it would be desirable to construct a much more rigorous analytic model for their reorientations from the initial tendency 
under the torque wrench effect. Although our analytic formula has been found to effectively describe the numerical results, it is not a physical model derived from the 
first principles.  We plan to work on this issue, hoping to report the result in the future. 

\acknowledgments

The IllustrisTNG simulations were undertaken with compute time awarded by the Gauss Centre for Supercomputing (GCS) 
under GCS Large-Scale Projects GCS-ILLU and GCS-DWAR on the GCS share of the supercomputer Hazel Hen at the High 
Performance Computing Center Stuttgart (HLRS), as well as on the machines of the Max Planck Computing and Data Facility 
(MPCDF) in Garching, Germany. We thank an anonymous referee for helping us significantly improve the original manuscript.  
JSM acknowledges the support by the National Research Foundation (NRF) of Korea grant funded by the Korean government (MEST) (No. 2019R1A6A1A10073437).
JL acknowledges the support by Basic Science Research Program through the NRF of Korea funded by the Ministry of Education (No.2019R1A2C1083855).

\clearpage
\clearpage
\clearpage
\clearpage
\clearpage
\end{document}